\documentclass{iopart}
\usepackage{times}
\usepackage{iopams}
\usepackage{latexsym}
\begin{document}

\title{Spin squeezing of mixed systems}

\author{K S Mallesh\dag\footnote[3]{\tt E-mail: ksmallesh@hotmail.com}, Swarnamala Sirsi\ddag
\footnote[4]{\tt E-mail: sshekar@bgl.vsnl.net.in}, Mahmoud A A Sbaih\dag,\\
P N Deepak\dag\footnote[5]{\tt E-mail: pndeepak@yahoo.com} and G Ramachandran\dag}

\address{\dag\ Department of Studies in Physics, University of Mysore,
 Mysore 570 006, India} 
\address{\ddag\ Department of Physics, Yuvaraja's College, University of Mysore, 
Mysore 570 005, India} 

\begin{abstract}
The notion of spin squeezing has been discussed in this paper using the density 
matrix formalism. Extending the definition of squeezing for pure states given by 
Kitagawa and Ueda in an appropriate manner and employing the spherical tensor 
representation, we show that mixed spin states which are non-oriented and possess 
vector polarization indeed exhibit squeezing. We construct a mixed state of a spin 
1 system using two spin 1/2 states and study its squeezing behaviour
as a function of the individual polarizations of the two spinors. 
\end{abstract}

\maketitle

\section{Introduction} 
The concept of spin is a fascinating topic in quantum theory.
Defined through the commutation relations ($\hbar =1$)
\begin{equation}
{\vec J}\times{\vec J}=i{\vec J},
\label{com}
\end{equation} 
which are common to intrinsic spin 
${\vec S}$ as well as orbital angular momentum ${\vec L}={\vec r} \times {\vec p}$,
it is interesting to note that equation (\ref{com}), in the case of $\vec L$ can
be derived from the position-momentum commutation relations 
\begin{equation}
\label{pm}
[k,p_{k}]=i\ ,\quad \,k=x,y,z. 
\end{equation} 
On the other hand, intrinsic spin associated with point particles like electrons 
are described in terms of the `up' spinor, $\it{u}$ and the `down' spinor, $\it{v}$
which are well-defined mathematically once the definition (\ref{com}) for spin is
accepted. Considering, therefore, the spinors as fundamental entities, 
Schwinger \cite{sch} visualized any state $\vert sm\rangle$ as made up of 
$s+m$ `up' spinors and $s-m$ `down' spinors through
\begin{equation}
\label{sch}
\vert sm\rangle={{(a^{\dagger}_{+})^{s+m}(a^{\dagger}_{-})^{s-m}}
\over {[(s+m)!(s-m)!]}^{1\over 2}}|00\rangle,
\end{equation}  
where $a^{\dagger}_{+},a^{\dagger}_{-}$ are the creation operators for the spin `up'
and spin `down' states, respectively.
According to Biedenharn and Louck \cite{bl}, this work of Schwinger was motivated by
the celebrated paper of Majorana \cite{maj}. It is well-known that the fundamental
uncertainty relation 
\begin{equation}
\label{unc}
\Delta x\,\Delta p_{x} \geq\,{1 \over 2},
\end{equation} 
of Heisenberg, which is equally valid for any pair of canonically 
conjugate variables, follows once (\ref{pm}) is postulated. Like wise, 
the uncertainty relations
\begin{equation}
\label{us}
\Delta S^{2}_{\alpha}\ \Delta S^{2}_{\beta} \geq \,
{1 \over 4}{\boldsymbol{\vert}}\langle \psi \vert S_{\gamma}\vert \psi \rangle
{\boldsymbol{\vert}}^{2},
\end{equation}  
for the spin operator $\vec S$, with $\alpha,\beta,\gamma=x,y,z$ cyclically,
are derivable for any spin state $\vert \psi \rangle$ once (\ref{com}) is postulated, although
no two components of $\vec S$ are canonically conjugate to each other.
It is also well-known that 
\begin{equation}
\Delta x\,=\Delta p_{x} =\,{1 \over \sqrt{2}},
\end{equation} 
in the case of the ground state of a simple harmonic oscillator in one dimension
and this corresponds to the minimum uncertainty given by the equality in (\ref{unc}). 
A state for which
\begin{equation}
\Delta x <{1 \over \sqrt{2}},
\end{equation} 
is then said to be squeezed in configuration space. One can similarly define a
squeezed state of the oscillator in momentum space. Just as the idea of coherent states
introduced by Schroedinger \cite{schr} for the harmonic oscillator was 
extended \cite{har,lit} to discuss
coherence in optics, the notion of squeezed states was also extended to the 
radiation field \cite{rad} and to spin states \cite{ku,puri,wine} as well.
A spin state may be said \cite{meh} to be squeezed if the variance $\Delta S_{\perp}$
associated with a spin component normal to the mean spin direction $\hat V$ satisfies
the condition,
\begin{equation}
\label{sc}
\Delta S^{2}_{\perp} < {1 \over 2}\left\vert\langle \psi 
\vert\vec {S}\cdot \hat {V} \vert \psi \rangle\right\vert .
\end{equation}  
A more stringent condition
\begin{equation}
\label{xi}
\xi=\left[{2s(\Delta S_{\perp})^{2} \over \left\vert \langle \psi \vert \vec S
\cdot \hat V\vert \psi \rangle\right\vert^{2}}\right]^{1\over 2}<1,
\end{equation}  
has been advocated by Wineland et al \cite{wine}.  Kitagawa and Ueda 
\cite{ku} have argued that it would be possible to cancel out
fluctuations in one direction normal to $\hat V$ at the expense of the other, 
provided quantum correlations are established among the elementary spinors 
which constitute a spin $s$ state in the sense of (\ref{sch}). Likewise a physical 
basis for (\ref{xi}) has also been discussed by Puri \cite{puri}. 
More recently \cite{meh}, a classification of pure states $\vert \psi \rangle$ into
two classes referred to as `oriented' and `non-oriented' has been
suggested employing a construction of states of spin $s$ out of $2s$
non-collinear spinors and it was explicitly shown in the case of $s=1$
that a state $\vert \psi \rangle$ has to be necessarily  non-oriented
for it to be a squeezed state.

The purpose of this paper is to extend the notion of spin squeezing to statistical 
assemblies of particles with spin $s$ as it will not only provide a complete spin
description of spin squeezing but also facilitate planning
of experiments to study squeezed spin states. This discussion is best 
done naturally by employing the language of the density matrix. 
An advantage of the density matrix formalism is that it can be applied with equal ease
to discuss pure as well as the mixed spin systems. This formalism is outlined in section 2
using the well known spherical tensor representation for the density matrix. 
In section 3 the squeezing condition (\ref{sc}) based on the uncertainty relation (\ref{us}) 
is generalized to take care of statistical assemblies as well.  
In section 4, we show that squeezing is exhibited by only non-oriented 
systems with non-zero vector polarization. In section 5 
we discuss the squeezing behaviour of a mixed 
spin 1 state which naturally arises in experiments \cite{expt}
employing polarized spin ${1\over2}$ beams on polarized spin ${1\over 2}$
targets. We also look into the spin-spin 
correlations which exist between these spinors when 
they are combined to yield a spin $1$ state.
    
\section{Density Matrix Description}
The density matrix $\rho$ for a spin $s$ system, pure or mixed has the
standard expansion
\begin{equation}
\label{rho}
\rho ={{{\rm Tr}\rho}\over{2s+1}}\sum_{k=0}^{2 s}
\sum_{q=-k}^{k}  (-1)^{q} t^{k}_{-q}\ \tau^{k}_{q},
\end{equation}
where $\tau^{k}_{q}$ (with $\tau^{0}_{0}
=I$, the identity operator) are irreducible tensor operators
of rank $k$ in the $n=2s+1$ dimensional spin
space with projection $q$ along the axis of quantization in the real 
3-dimensional space.  The $\tau^{k}_{q}$ satisfy the commutation relations
\begin{eqnarray}
\fl\left[\tau^{k_{1}}_{q_{1}},\tau^{k_{2}}_{q_{2}}\right]=[s][k_{1}][k_{2}]\sum^{k_{1}+k_{2}}_{k=\vert k_{1}-k_{2}
\vert} \left(1-(-1)^
{k_{1}+k_{2}-k}\right)\ C(k_{1}k_{2}k;q_{1}q_{2}q)\nonumber\\ 
\lo \times
W(sk_{1}sk_{2};sk)\ \tau^{k}_{q}
\end{eqnarray}
where $C$ and $W$ denote Clebsch-Gordan and Racah coefficients respectively and 
we use the short hand $[s]=\sqrt{2 s+1}$.
They also satisfy the orthogonality relations
\begin{equation}
{\rm Tr}\{\tau^{k^{\dagger}}_{q}\tau^{k'}_{q'}\}=n\ \delta_{kk'}\delta_{qq'}.
\end{equation}
Here the normalization has been chosen so as to be in agreement
with Madison convention \cite{mad}. The Fano statistical tensors 
or the spherical tensor parameters $t^{k}_{q}$ in (\ref{rho}) which characterize
the given system are the average expectation values given by    
\begin {equation} 
t^{k}_{q}={\rm Tr} \{ \rho\tau^{k}_{q}\}/{\rm Tr}\rho. 
\end{equation}
Since $\rho$ is Hermitian, and $\tau^{k^{\dagger}}_{q}=(-1)^{q} \tau^{k}_{-q}$, 
these satisfy the condition
\begin{equation}
t^{k^{\star}}_{q} =(-1)^{q}t^{k}_{-q}. 
\end{equation}
Apart from $t^{0}_{0}={\rm Tr} \rho$, there are $n^{2}-1=4s(s+1)$
real independent parameters for the most general mixed state.
The density matrix $\rho$ for pure state satisfies $\rho^{2}=\rho$,
and hence a normalized pure state has only $4s$ real independent
parameters describing it.  This leads to a set of constraints
\begin{equation}
[k]\sum _{k_{1},k_{2}} [k_{1}][k_{2}]\ W(sk_{1} sk_{2};sk)\ (t^{k_{1}}
\otimes t^{k_{2}})^{k}_{q}=[s]\ t^{k}_{q}
\end{equation}
on $t^{k}_{q}$ for each $k$ and $q$.
It is worth noting here that in addition to the above representation
for density matrix which uses spherical tensor operators, there
also exist other representations such as the $SU(n)$ representation \cite{mvn},
where the density matrix is expanded in terms of the generators
of the Lie group $SU(n)$ ,whose number is also $n^{2}-1$.
This representation is advantageous since the diagonal form of 
$\rho$ can be expressed in 
terms of the subset consisting of diagonal generators which are 
$n-1$ in number.  On the other hand, 
the spherical tensor representation which is widely used in spin physics
has the advantage that the spherical tensor parameters have
simple transformation properties under coordinate rotations in the
real 3-dimensional space.  If a coordinate frame I is transformed to II 
through  a rotation $R(\alpha,\beta,\gamma)$,where  $\alpha,\beta,\gamma$
are the Eulerian angles, the $t^{k}_{q}$ in the respective
frames are related through
\begin {equation}
(t^{k}_{q})_{II}=\sum _{q'} D^{k}_{q'q}( \alpha,\beta,\gamma)(t^{k}_{q'})_{I},
\end {equation}
where $D^{k}_{q'q}(\alpha,\beta,\gamma)$ is the matrix representation
of the rotation. The spherical tensor operators $\tau ^{k}_{q}$
are homogeneous polynomials of rank $k$ and projection $q$, constructed
out of the spin operators $S_{x}, S_{y}$ and $S_{z}$. In particular, 
the operator
$\vec S$ is a vector (rank 1) operator and its spherical components
are related to $\tau^{1}_{q}$ through
\begin {equation}
S^{1}_{q}=\left[{{s(s+1)} \over 3}\right]^{1\over 2}\ \tau^{1}_{q}\ ;\ q=1,0,-1.
\end {equation}
The average expectation value of $\vec S$ in the state specified by $\rho$
given by
\begin{equation}
\label{P}
{\vec P}={{{\rm Tr}\{\rho {\vec S}\}} \over {\rm Tr}\rho}
\end{equation}
is called the vector polarization in spin physics literature.
Kitagawa and Ueda \cite{ku} refer to this as the 
{\it mean spin vector} in their paper.
The spherical components of $\vec P$,
\begin{equation}
P_{\pm 1}=\mp{1\over {\sqrt 2}}(P_{x}\pm iP_{y})\ ;\ P_{0}=P_{z},
\end{equation}
are related to $t^{1}_{q}$ through
\begin{equation}
P^{1}_{q}={{{\rm Tr}\{\rho S^{1}_{q} \}} \over {{\rm Tr}\rho}}=
\left[{{s(s+1)} \over 3}\right]^{1\over 2}\ {{t^{1}_{q}}}.
\end {equation}
The expectation values of other observables such as $S_{x}^{2}, S_{y}^{2},S_{z}^{2}$
on the other hand are related to the alignment parameters $t^{2}_{q}$
through
\begin{equation}
{\rm Tr}\{\rho S_{x}^{2}\}={\textstyle{1\over {f_{1}^{2}}}} -{\textstyle{1\over {f_{2}\sqrt 6}}}\ t^{2}_{0}
+{\textstyle{1\over {2f_{2}}}}(t^{2}_{2}+t^{2}_{-2})
\end {equation}
\begin{equation}
{\rm Tr}\{\rho S_{y}^{2}\}={\textstyle{1\over {f_{1}^{2}}}} -{\textstyle{1\over {f_{2}\sqrt 6}}}\ t^{2}_{0}
-{\textstyle{1\over {2f_{2}}}}(t^{2}_{2}+t^{2}_{-2})
\end {equation}
\begin{equation}
{\rm Tr}\{\rho S_{z}^{2}\}={\textstyle{1\over {f_{1}^{2}}}} +{\textstyle{1\over f_{2}}}{\textstyle\sqrt {2\over 3}}\ t^{2}_{0},
\end {equation}
where
\begin{equation}
f_{1}=\left[{3 \over {s(s+1)}}\right]^{1\over 2}\ ;\ f_{2}=
\left[{30 \over {s(s+1)(2s-1)(2s+3)}}\right]^{1\over 2}.
\end {equation}
These lead to the variances
\begin{eqnarray}
\fl\Delta S_{x}^{2}&=&{{{\rm Tr}\{\rho S_{x}^{2}\}}\over {{\rm Tr}\rho}} 
-\left[{{{\rm Tr}\{\rho S_{x}\}}\over {{\rm Tr}\rho}}\right]^{2}\nonumber \\
\fl &=&{1\over {{\rm Tr}\rho}}\left[{\textstyle{1\over {f_{1}^{2}}}} -{\textstyle{1\over {\sqrt 6}f_{2}}}t^{2}_{0}+
{\textstyle{1\over {2f_{2}}}}(t^{2}_{2}+t^{2}_{-2})\right]-
{\textstyle{1\over {2f_{1}^{2}}}}{1\over {({\rm Tr}\rho)^{2}}}(t^{1}_{-1}-t^{1}_{1})^{2}
\end{eqnarray}
\begin{eqnarray}
\fl\Delta S_{y}^{2}&=&{{{\rm Tr}\{\rho S_{y}^{2}\}}\over {{\rm Tr}\rho}} 
-\left[{{{\rm Tr}\{\rho S_{y}\}}\over {{\rm Tr}\rho}}\right]^{2}\nonumber \\
\fl &=&
{1\over {{\rm Tr}\rho}}\left[{\textstyle{1\over {f_{1}^{2}}}} -{\textstyle{1\over {\sqrt 6}f_{2}}}t^{2}_{0}-
{\textstyle{1\over {2f_{2}}}}(t^{2}_{2}+t^{2}_{-2})\right]+
{\textstyle{1\over {2f_{1}^{2}}}}{1\over {({\rm Tr}\rho)^{2}}}(t^{1}_{-1}+t^{1}_{1})^{2}
\end{eqnarray}
\begin{eqnarray}
\fl\Delta S_{z}^{2}&=&{{{\rm Tr}\{\rho S_{z}^{2}\}}\over {{\rm Tr}\rho}} 
-\left[{{{\rm Tr}\{\rho S_{z}\}}\over {{\rm Tr}\rho}}\right]^{2}\nonumber \\
\fl &=&{1\over {{\rm Tr}\rho}}\left[{\textstyle{1\over {f_{1}^{2}}}} - 
{\textstyle{1\over {f_{2}}}}{\textstyle{\sqrt {2\over 3}}}t^{2}_{0}\right]-
{\textstyle{1\over {f_{1}^{2}}}}{1\over {({\rm Tr}\rho})^{2}}(t^{1}_{0})^{2}.
\end{eqnarray}
While a system in a pure state satisfying $\rho^{2}=\rho$ is completely polarized,
a system in a mixed state is either partially polarized or unpolarized.
For an unpolarized system, $t^{k}_{q} = 0$ for all $k= 1,\ldots,2s$. A partially
polarized or completely polarized state is said to be vector polarized
if ${\vec P}\neq 0$ and aligned or tensor polarized if at least
one $t^{2}_{q} \neq 0$.   A Cartesian coordinate frame chosen with its ${\hat z}-$axis
parallel to ${\vec P}$ is referred to as Lakin Frame [LF] \cite{lf}.  
In other words, in such 
a frame $t^{1}_{\pm 1}=0$.  On the other hand, a frame in which $t^{2}_{2}$ 
is real and $t^{2}_{\pm 1}=0$ is referred to as the Principal Axes of Alignment
Frame (PAAF) \cite{paaf}.  The latter is also identified as a frame 
in which the traceless second rank 
Cartesian tensor $P_{\alpha\beta}$ (which is 
defined by the $t^{k}_{q}$) is diagonal. 
While there is only one PAAF for a system up to possible renaming of the axes, 
Lakin frame on the other hand requires only ${\hat z}_{_{0}}$ axis to be along ${\hat P}$
and depending on the choice of $x_{_{0}}$ and $y_{_{0}}$ axes, we have an 
infinite number of LFs.
 
\section{Spin squeezing}
The Heisenberg uncertainty relationship for the spin operators
$S_{x}, S_{y}$ and $S_{z}$ satisfying (\ref{com}) is given by (\ref{us}),
where the variance $\Delta S^{2}_{i}$ and the average expectation 
value $\langle S_{z}\rangle$
depend not only on the spin state but also
on the frame with respect to which the spin operators have been specified.
Following Kitagawa and Ueda \cite{ku} and Puri \cite{puri}, we have 
defined squeezing criterion in our earlier
paper \cite{meh} as follows.  Given the quantum state 
$\vert\psi\rangle$
of spin $s$, the mean spin direction associated with it is 
given by
\begin{equation}
{\hat P}={\frac{\langle\psi\vert{\vec S}\vert\psi\rangle}
{\left\vert\langle\psi\vert{\vec S}\vert\psi\rangle\right\vert}}.
\end{equation}
The state $\vert\psi\rangle$ is then said to be squeezed in the spin component
$S_{{}_{\perp}}={\vec S}\cdot{\hat P}_{{}_{\perp}}$, if
\begin{equation}
\label{sper}
(\Delta S_{{}_{\perp}})^{2}< {1\over 2}\left\vert\langle{\vec S}\cdot{\hat P}
\rangle\right\vert,
\end{equation}
where ${\hat P}_{{}_{\perp}}$ is orthogonal to ${\hat P}$.   This criterion of 
squeezing aims at characterizing squeezing as an intrinsic feature
of the state and Kitagawa and Ueda \cite{ku} have remarked that if the spin state 
is visualized as being made up of $2s$ spin $1\over 2$  
states, then the quantum  correlations that exist among these component
spins are responsible for the manifestation of squeezing in the given
quantum state. They substantiate their statement through a pictorial
representation in which they show that a spin coherent state which is
built out of $2s$ spinors all oriented in the same direction, is not 
squeezed as there exist no quantum correlations in such an arrangement.
On the other hand, a squeezed state of spin is depicted as being built out
of the same number of the spins which possess quantum correlation.
We have looked into this aspect, in all its details, in the case of spin 1
in our earlier paper \cite{meh} and an explicit connection between the spin-spin
correlations and spin squeezing has been shown to exist. In the light of this
we now adopt for the case of mixed states the generalized  form of the above
criterion.  Explicitly, a spin state specified by $\rho$ is said to be squeezed
in the component $S_{{}_{\perp}}(\equiv{\vec S}\cdot{\hat P}_{{}_{\perp}})$, if
\begin{equation}
\label{sq}
\Delta({\vec S}\cdot{\hat P}_{{}_{\perp}})
^{2}={\frac {{\rm Tr}\{\rho({\vec S}\cdot{\hat P}_{{}_{\perp}})^{2}\}}
{{\rm Tr}\rho}}<
{1\over 2}\left\vert\langle{\vec S}\cdot{\hat P}\rangle\right\vert=
{\frac {{\boldsymbol\vert} {\rm Tr}\{
\rho{\vec S}\cdot{\hat P}\}{\boldsymbol\vert}}{2 { \rm Tr}\rho}}
\end{equation}
where ${\hat P}_{{}_{\perp}}$ denotes any direction which is orthogonal to 
the vector polarization ${\vec P}$.
It may be noted here that the squeezing criterion defined here is 
distinct from several others used in the literature \cite{lit}.  For example, 
if one uses the criterion
\begin{equation}
\Delta S_{i}^{2}< {1\over 2}{\boldsymbol\vert}\langle S_{z}\rangle{\boldsymbol\vert}\  ;\  i=x,y,
\end{equation}
where the components are referred to a frame chosen arbitrarily
then, as has been pointed out by Kitagawa and Ueda \cite{ku}, it turns out  that
a given state will be squeezed with respect to a component in one frame but will not be 
so in another frame.  This aspect makes squeezing solely frame-dependent
and extrinsic to the system.  On the other hand, the above form (\ref{sq}) of the 
criterion suggests that a quantum state itself 
specifies a direction with respect to which it reveals the presence of squeezing
in its spin component. As such, this criterion, which we adopt here,
characterizes any manifestation of squeezing as an intrinsic 
property of a spin system, like in the case of radiation field.
We now classify, as in our earlier paper, the spin states into
oriented and non-oriented states and study the squeezing aspect
of states in each class based on this criterion (\ref{sq}).

\section{Mixed state classification and squeezing}
\subsection {Oriented system}
A spin system is said to be oriented \cite{ori} if
its density matrix $\rho$ has a diagonal form
$\rho_{_{0}}$ with its eigen states being the angular momentum basis 
states $\vert s m\rangle _{_{0}}$
referred to the axis of quantization ${\hat z}_{_{0}}$. In other words, an oriented 
system is one in which the populations are distributed with respect
to the basis states $\vert sm\rangle _{_{0}}$ and $\hat z_{_{0}}$ is then 
called the axis of 
orientation.  It may be noted here that this definition for mixed 
states is a natural generalization of the definition of an oriented
pure state defined in our earlier paper \cite{meh}.  If $p_{m}$ denote the fractional
populations of an oriented system in the states $\vert sm\rangle _{_{0}}$ , the 
density matrix $\rho_{{}_{0}}$ is given by the expansion
\begin{equation}  						   
\rho _{{}_{0}}=\sum p_{m}\vert sm\rangle_{_{0}}\ {}_{_{0}}\langle sm\vert,
\ p_{m}\geq 0\ ;\ \sum p_{m}=1.
\end{equation}
The vector polarization ${\vec P}$, for such a system turns out to be 
along ${\hat z}_{_{0}}$ itself and in a LF whose ${\hat z}-$axis
is along ${\hat z}_{_{0}}$, we have 
\begin{equation} 
{\vec P}=\left(0, 0,\sum_{m} m\ p_{m}\right)=\left(\sum_{m} m\ p_{m}
\right){\hat z}_{_{0}}.
\end{equation}
Any vector ${\hat P}_{{}_{\perp}}$ perpendicular to ${\hat P}$, therefore
lies in the $xy$-plane of the chosen LF and we can express it as
\begin{equation}   
{\hat P}_{{}_{\perp}}={\hat x}\cos\phi+{\hat y}\sin\phi\ ;\ 0\leq\phi<2\pi.
\end{equation}
This makes
\begin{equation}  
{\vec S}\cdot{\hat P}_{{}_{\perp}}=S_{x}\cos\phi + S_{y}\sin\phi
\end{equation}
and we have 
\begin{equation}
\langle{\vec S}\cdot{\hat P}_{{}_{\perp}}\rangle=
\langle S_{x}\rangle \cos\phi 
+\langle S_{y}\rangle \sin\phi=0
\end{equation}
since $\langle S_{x}\rangle=\langle S_{y}\rangle=0$ in the Lakin frame.
The variance in ${\vec S}\cdot {\hat P}_{{}_{\perp}}$ then becomes
\begin{equation}
\Delta ({\vec S}\cdot{\hat P}_{{}_{\perp}})^{2}={1\over 2}\left
[s(s+1)-\sum_{m} m^{2} p_{m}\right],
\end{equation}
so that the squeezing criterion (\ref{sq}) for an oriented system takes the form
\begin{equation}
s(s+1)-\sum_{m} m^{2}\  p_{m}<\left\vert\sum_{m} m\  p_{m}\right\vert,
\end{equation}
and it is quite easy to see that this inequality is never satisfied 
for any value of $s$.   Thus we arrive at the conclusion that no oriented
system, either pure or mixed, is squeezed . It is interesting 
to note here that every spin $1\over 2$ state, either pure or mixed, is 
always oriented. This is due to the property that any spin $1\over 2$ density
matrix can always be diagonalized by an appropriate unitary matrix
belonging to the group $SU(2)$, the latter bearing the property that it provides 
a representation of the rotation group in 3-dimension.  In other 
words the eigen states of the density matrix $\rho$ for a spin $1\over 2$
system can always be identified as the spin -up and spin down 
states with respect to an appropriate axis of quantization.  This, together
with what has been said above, implies that squeezing is absent 
in the case of spin $1\over 2$, irrespective of whether the  state
is pure or mixed.
We now look at the second class of spin states namely the 
{\it non-oriented} spin systems and look at their squeezing behaviour
in what follows.
  
\subsection {Non-oriented system (states)}
A non-oriented spin $s$ state $\vert\psi\rangle$ has been defined
earlier \cite{mvn} as one which can not be identified as eigen
state of $S_{z}$ with any choice of ${\hat z}-$axis as the axis
of quantization.  We may therefore define a mixed non-oriented 
system as one where the eigen states 
$\vert\psi_{1}\rangle,\vert\psi_{2}\rangle,\ldots,\vert\psi_{n}\rangle$
of the density matrix $\rho$ can not all be identified with states
$\vert sm\rangle,\ m=-s,\ldots,s$ with respect to any suitable
${\hat z}-$axis.  In other words, at least one of the eigen states
$\vert\psi_{i}\rangle,\ i=1,\ldots,n$  has to be non-oriented as defined
earlier.  The system will be maximally non-oriented if every
one of the eigen states is non-oriented.  Such non-oriented
systems can exist only for spin $s\geq 1$ since the unitary group in
$n$-dimension is homomorphic to the rotation group in 3-dimensions
only in the particular case of $n =2$.  

While an oriented system gets specified through the distribution
of populations in angular momentum states with respect to a single axis namely
the axis of orientation, it has been shown by Ramachandran and Ravishankar \cite{vr}
that a non-oriented system is characterised by more than one axis.
This identification has been arrived at using the spherical tensor 
representation for the density matrix of such a system.  In the most general case 
it has been shown by them that a set of $s(2s+1)$ axes ${\hat Q}_{i}, i = 
1,\ldots,s(2s+1)$
are needed to characterize a non-oriented state.  Each $t^{k}_{q}$
in (\ref{rho}) for any given $\rho$ can be expressed as     
\begin{equation}
t^{k}_{q}=r_{k}\Bigl(\ldots(Y_{1}({\hat Q}_{1})\otimes Y_{1}({\hat Q}_{2}))^{2}\otimes
\ldots )^{k-1}\otimes Y_{1}({\hat Q}_{k})\Bigr)^{k}_{q},
\end{equation}
where $r_{k}$ is a real constant and $Y_{1m}({\hat Q}_{i})$ is the 
spherical harmonic function associated with the direction ${\hat Q}_{i}$.
An oriented system, in this language, is one for which all the $s(2s+1)$
axes merge together to give a single axis ${\hat Q}_{0}$, which is itself
the axis of orientation. The choice of ${\hat z}-$axis along ${\hat Q}_{0}$
for an oriented system leads to the vanishing of all $t^{k}_{q}$
with $q \neq 0$ and therefore, an oriented
system in its LF is described by the Fano statistical tensors $t^{k}_{0}$
only. The non-oriented systems, on the other hand, possess, 
in general, non-zero $t^{k}_{q}$ with respect to any angular momentum
basis.

Before we look into the aspect of squeezing, it may be appropriate to
briefly mention the nomenclature for the specific kinds of spin
systems, often adopted in the spin physics literature.  A spin system
with non-zero $t^{1}_{q}$ is said to be vector polarized while that
with non-zero $t^{2}_{q}$ is said to be aligned. A purely aligned
system has non-zero $t^{2}_{q}$ but all other tensor polarizations
including the vector polarization $t^{1}_{q}$ will be zero.

Coming back to the notion of squeezing, it is to be noted that for
the squeezing criterion to be satisfied, the system should necessarily
possess non-zero vector polarization since only then the right hand
side of the inequality in (\ref{sq}) will be non-zero and one can look for the
satisfiability of the squeezing criterion.  If the vector polarization
is zero then every frame qualifies to be a Lakin frame and since
$\Delta ({\vec S}\cdot{\hat i})^{2}$ is always non-negative for any direction
$\hat i$, we conclude that all non-oriented states which do not possess
vector polarization lack squeezing. One can, however, define
higher order squeezing behaviour via a proper criterion
and examine such situations.
Having ruled out squeezing in the case of oriented systems and in the case 
of non-oriented systems with zero vector polarization, we are left with
non-oriented systems which possess non-zero vector polarization.  Let
us suppose that the density matrix of such a system is specified
with respect to the angular momentum basis $\vert s m \rangle$ relative
to a frame $xyz$ in terms of the spherical tensors $t^{k}_{q}$ through
(\ref{rho}).  We now make a transition to a particular LF $x_{_{0}}y_{_{0}}z_{_{0}}$
in the following way. The vector polarization direction ${\hat P}={\hat z}_{_{0}}$
of the system is determined by using (\ref{P}).  If $(\theta _{0},\phi _{0})$
denote the direction of ${\hat z}_{_{0}}$ with respect to $xyz$, the 
frame $xyz$ is rotated first about
the ${\hat z}-$axis through $\phi _{0}$ and then about the new ${\hat y}-$axis through
$\theta _{0}$. The frame so obtained (call it $x'y'z_{_{0}}$) is a Lakin frame 
as the ${\hat z}-$axis of $xyz$ now coincides with ${\hat z}_{_{0}}$.  The spherical 
tensor parameters $t^{k}_{q}$ that specify the density matrix
$\rho$ in this frame are related to $t'^{k}_{q}$ through 
\begin{equation}
t^{k}_{q}=\sum D^{k}_{q'q}(\phi _{0},\theta _{0},0)\ t'^{k}_{q} .
\end{equation}
While this frame is
enough for the purpose of identifying squeezing, we wish to use the additional
degree of freedom of rotating $x'y'z_{_{0}}$ about ${\hat z}_{_{0}}$ through an
angle $\gamma$ to get a special LF.  The angle $\gamma$
here is so chosen that the second rank tensor $t^{2}_{2}$ after 
the rotation is real.  With this choice of
$x_{_{0}} y_{_{0}} z_{_{0}}$, $t^{1}_{1}=t^{1}_{-1}=0$
and $t^{2}_{2}=t^{2}_{-2}$, the first
being due to transition to a LF while the second
being due to the use of additional degree of freedom of rotation
about ${\hat z}_{_{0}}$ through $\alpha$. In the special LF
$x_{_{0}} y_{_{0}} z_{_{0}}$, we then collect the relevant quantities
needed for identifying squeezing given by
\begin{equation}
\langle S_{z_{0}}\rangle ={1\over {f_{1}}}t^{1}_{0} \quad ;
\quad \langle S_{x_{0}}\rangle=\langle S_{y_{0}}\rangle=0 
\end{equation}
\begin{equation}
\Delta S_{x_{0}}^{2}={1\over {f_{1}^{2}}}+{1\over {2f_{2}}}
\left(2t^{2}_{2}-{\textstyle{\sqrt {2\over 3}}}
t^{2}_{0}\right)
\end{equation}
\begin{equation}
\Delta S_{y_{0}}^{2}={1\over {f_{1}^{2}}}-
{1\over {2f_{2}}}\left(2t^{2}_{2}+ {\textstyle{\sqrt {2\over 3}}}
t^{2}_{0}\right).
\end{equation}
Defining $S_{\perp}$ as in equation (\ref{sper}) we get 
\begin{equation}
\Delta S^{2}_{\perp}=\Delta S^{2}_{x_{0}}\cos^{2}\phi +\Delta S^{2}_{y_{0}}\sin^{2}\phi
\end{equation}
so that the squeezing criterion for $S_{\perp}$ takes the form 
\begin{equation}
\fl 1+\left[{{3(2s+3)(2s-1)}\over{s(s+1)40}}\right]^{1\over 2}\left(2 t
^{2}_{2}\cos 2\phi - {\textstyle{\sqrt {2\over 3}}}t^{2}_{0}\right)
\ <\ {1\over 2} \left[{3 \over {s(s+1)
}}\right]^{1\over 2} {\boldsymbol\vert} t^{1}_{0}{\boldsymbol\vert} ,
\end{equation}  
for any value of $\phi,\ 0\leq \phi\leq 2\pi$.
States satisfying this criterion are then squeezed for those $\phi,\ 0\leq \phi\leq 2\pi$
in the component of spin. 
In specific cases, this inequality is indeed
satisfied over a range of values for $t^{k}_{q}$ and we
therefore conclude that squeezing is indeed exhibited by only
non-orenited states with non-zero vector polarization.
To support the claim made here we present below in table (1) several situations
which reveal the presence of squeezing. 

We wish to note here that $t^{k}_{q}$ present in the table above
actually correspond to realistic situations as they have been chosen 
in accordance with the positive semi-definiteness requirement
of the density matrix $\rho$.  For example, for $s=1$, this property of 
$\rho$ implies that the spherical tensor parameters have to satisfy
the boundary conditions \cite{min}
\begin{equation}
0 \le {1\over 3}(1\pm{\textstyle{\sqrt {3 \over 2}}}\ t^{1}_{0}+ 
{\textstyle{1\over {\sqrt 2}}}
\ t^{2}_{0})\le 1
\end{equation}
\begin{equation}
0\le {1\over 3}(1-{\sqrt 2}\ t^{2}_{0})\le 1
\end{equation}
\begin{equation}
0\le (t^{1}_{0})^{2}+2|t^{2}_{2}|^{2}+2 
{\boldsymbol\vert} t^{2}_{1}{\boldsymbol\vert} ^{2}+
(t^{2}_{0})^{2}\le 2
\end{equation}
 \begin{equation}
 0 \le\ {\rm det}\ \rho  \le {1 \over 27}
\end{equation}
If all the $t^{k}_{q}$s are treated as the component of a $(2s+1)^{2}-1$ 
dimensional complex vector ${\vec T}$, then, when the spin system is subjected 
to an interaction, this vector ${\vec T}$ starts moving in the $(2s+1)^{2}-1 $ 
dimensional complex space, of course, subjected to the above constraints.
It is therefore natural to ask how the squeezing evolves during such an
evolution. This behaviour merits an independent study which is being taken
up at present. 

\section{Squeezing of channel spin 1 states} 
The concept of channel spin plays an important role in hadron scattering 
and reaction processes.  Consider for example, a beam of nucleons colliding
with a proton target both of which are prepared initially to be mixed states
specified by their density matrices 
 \begin{equation}
 \rho(i)= {1 \over 2}\big[1+ {\vec \sigma}(i)\cdot{\vec P}(i)\bigr]= {1 \over 2}
 \sum_{k,q} t^{k}_{q}(i)\tau^{k^{\dagger}}_{q}(i)\  ;\  i=1,2.
 \end{equation}
 Channel spin states $s=0,1$ come into play in scattering and reaction 
 process \cite{pnd}.
The combined density matrix $\rho_{c}$ is the direct product 
 of these two density matrices, i.e.,
 \begin{equation}
 \rho_{c}=\rho_{1}\otimes \rho_{2},
 \end{equation}
 and the density matrix for the channel spin 1 state is given by 
 \begin{equation}
 \rho = \left[{{3+{\vec P}(1)\cdot{\vec P}(2)}\over {12}}\right]\left[1
 + \sum _{k,q}t^{k}_{q} \tau ^{k^{\dagger}}_{q}\right],
\end{equation}
where the spherical tensor parameters $t^{k}_{q}$ are related to 
the individual $t^{k}_{q}(i)$ through
\begin{equation}
\label{cou}
\fl t^{k}_{q} = {\rm Tr}(\rho\ \tau^{k}_{q})=\left[{{6{\sqrt 3}}\over {3+
{\vec P}(1)\cdot{\vec P}(2)}}\right]\sum _{k_{1},k_{2}}[k_{1}][k_{2}]
\left\{
\matrix{
{\textstyle{1\over 2}} & {\textstyle{1\over 2}} & k_{1}\cr
{\textstyle{1\over 2}}&{\textstyle{1\over 2}}& k_{2}\cr
1& 1& k\cr
}
\right\}
\left(t^{k_{1}}(1)\otimes t^{k_{2}}(2)\right)^{k}_{q}.
\end{equation}
In equation (\ref{cou}), $\{\}$ denotes the Wigner 9-$j$ symbol \cite{rus}.
Explicitly we have
\begin{equation}
\label{t1}
t^{1}_{q} =\left[{{\sqrt 6} \over{3+{\vec P}(1)\cdot{\vec P}(2))}}\right]
({\vec P}_{q}(1) +{\vec P}_{q}(2))
\end{equation}
\begin{equation}
\label{t2}
t^{2}_{q} =\left[{{2{\sqrt 3}} \over{3+{\vec P}(1)\cdot{\vec P}(2))}}\right]
({\vec P}(1)\otimes{\vec P}(2))^{2}_{q}.
\end{equation} 
In order to discuss the squeezing nature of the channel state, we have
to first choose a Lakin frame. A glance at the above equation for
$t^{1}_{q}$ suggests that the ${\hat z}_{_{0}}-$axis (of the LF)
should be chosen along ${\vec P}(1)+{\vec P}(2)$. Since $
{\vec P}(1),{\vec P}(2)$ together define a plane in any general situation, 
we choose ${\hat x}_{_{0}}$ axis to be in
this plane such that the azimuths of ${\vec P}(1),{\vec P}(2)$
with respect to ${\hat x}_{_{0}}$ are respectively 0 and $\pi$. The $y_{_{0}}$ axis is
then chosen to be along ${\hat z}_{_{0}}\times {\hat x}_{_{0}}$.  The frame so chosen
is indeed the Special LF (see figure 1)
\begin{figure}
\begin{center}
\caption{Special Lakin Frame $x_{_{0}}y_{_{0}}z_{_{0}}$, where
${\hat z}_{_{0}}$ is along ${\vec P}(1)+{\vec P}(2)$, ${\hat x}_{_{0}}-$
axis in the plane of ${\vec P}(1),{\vec P}(2)$ such that 
the azimuths of ${\vec P}(1),{\vec P}(2)$ are $0,\pi$
respectively.}
\end{center}
\end{figure}
as is evident from 
equations (\ref{t1}) and (\ref{t2}) that $t^{1}_{\pm 1}=0$ and 
$t^{2}_{2}=t^{2}_{-2}$.  In this 
frame so chosen, we have   
\begin{equation}    
\fl P_{x_{0}}(1)={\frac {P(1)P(2)\sin\theta}{|{\vec P}(1)+{\vec P}(2)|}} = - P_{x_{0}}(2)
\end{equation}				 
\begin{equation}
\fl P_{y_{0}}(1) =P_{y_{0}}(2)=0
\end{equation}	  
\begin{equation}  
\fl P_{z_{0}}(1)={\frac {P(1)^{2}+P(1)P(2)\cos\theta}{|{\vec P}(1)+{\vec P}(2)|}}\ ;\
P_{z_{0}}(2)={\frac {P(2)^{2}+P(1)P(2)\cos\theta}{|{\vec P}(1)+{\vec P}(2)|}}.
\end{equation}
If now $S_{\perp}$ is defined as $S_{x_{0}}\cos\phi+S_{y_{0}}\sin\phi$, then
the variance $\Delta S_{\perp}^{2}$ takes the form
\begin{equation}
\Delta S_{\perp}^{2}={{2[|{\vec P}(1)+{\vec P}(2)|^{2}-P(1)^{2}P(2)^{2}\sin^{2}\theta\cos^{2}\phi
]} \over ({3+{\vec P}(1)\cdot{\vec P}(2))|{\vec P}(1)+{\vec P}(2)|^{2}}},
\end{equation}
while the expectation value of $S_{z_{0}}$ will be given by 
\begin{equation}
\langle S_{z_{0}}\rangle={\frac{2|{\vec P}(1)+{\vec P}(2)|}{
(3+{\vec P}(1)\cdot{\vec P}(2))}}.
\end{equation}
The squeezing condition for $S_{\perp}$ then becomes
\begin{equation}
1-
{\frac{\left\vert{\vec P}(1)\times{\vec P}(2)\right\vert^{2}}
{\left\vert{\vec P}(1)+{\vec P}(2)\right\vert^{2}}}
\ \cos^{2}\phi<{1\over 2}|{\vec P}(1)+{\vec P}(2)|.
\end{equation}
This expression has been studied numerically for several cases 
of ${\vec P}(1),{\vec P}(2)$
and $\phi$.  Squeezing is seen for a wide range of values of ${\vec P}(1)$, 
${\vec P}(2)$ and $\phi$ and in particular, 
maximum squeezing occurs when $\phi=0$ for any given ${\vec P}(1),{\vec P}(2)$.  
In other words, it is the spin
component $S_{x_{0}}$ of the Special LF which is maximally
squeezed.  A plot of the quantity 
\begin{equation}
\fl Q={1\over 2}|\langle S_{z_{0}}\rangle|-\Delta S^{2}_{x_{0}}={1\over 2}|
{\vec P}(1)+{\vec P}(2)|+
{\frac{\left\vert{\vec P}(1)\times{\vec P}(2)\right\vert^{2}}
{\left\vert{\vec P}(1)+{\vec P}(2)\right\vert^{2}}}
\ \cos^{2}\phi-1
\end{equation}
as a function of the angle $\theta$ between the two polarization
vectors reveals that the component $S_{x_{0}}$ is squeezed over a 
wide range of $\theta$ as is evident from the figure (2) and figure (3) shown below.
The graphical study also reveals that squeezing appears only when 
the degree of polarization of both the spins are more than 77\% of that
for a pure state in each case.  In particular, when the states are pure, 
the combined system will also be in a pure state but the spin 1 projection
of this pure state will be in an entangled state (refer to equation 
(25) in reference \cite{meh}).
\begin{figure}
\begin{center}
\caption{Variation of squeezing in $S_{x_{_{0}}}$ with respect to $\theta,$ the
angle between the two polarization vectors ${\vec P}(1)$ and ${\vec P}(2)$.}
\end{center}
\end{figure}
\begin{figure}
\begin{center}
\caption{Variation of squeezing in $S_{\perp}$ with respect to $\theta,$ the
angle between the two polarization vectors ${\vec P}(1)$ and ${\vec P}(2)$.}
\end{center} 
\end{figure} 
In this state, the 
squeezing condition reduces to 
\begin{equation}
\cos^{2}2\theta<|\cos 2\theta|
\end{equation}    
which agrees with the result obtained in our earlier paper \cite{meh}
(except that we have called the angle between ${\vec P}(1)$ and 
${\vec P}(2)$ as $2\theta$ here, while it is taken as $\theta$ there).
The origin of the squeezing behaviour of the coupled spin 1 system can be traced
as arising due to the intrinsic quantum correlations that exist between the individual 
spinors.  These correlations can be classified as (1) those that arise due to 
the coupling of the two subsystems and (2) those that arise when the combined
total density matrix $\rho _{_{C}}$ is projected on to the desired spin space.
In our present case, we have taken $\rho _{_{C}}$ to be a direct product of the 
two subsystem density matrices $\rho (1)$ and $\rho (2)$.  Such a $\rho _{_{C}}$
is not entangled and therefore there are no correlations of the first kind.
However, when we take the spin 1 projection of $\rho _{_{C}}$, the correlations
of the second type will appear in the spin 1 projection.  These correlations
are given by
\begin{eqnarray}
\fl C^{12}_{xx}&=&{\frac{P^{2}_{s}-P_{d}(P(1)^{2}+P(2)^{2})
-2P(1)^{2}P(2)^{2}(1+\sin^{2}\theta\cos 2\phi)}{ 4 (3+P_{d})P^{2}_{s}}}\\
\fl C^{12}_{yy}&=&{
 { P^{2}_{s}}-{2P(1)^{2}P(2)^{2}(1-\sin^{2}\theta\cos 2\phi)}
-{ P_{d}(P(1)^{2}+P(2)^{2})}
\over {4}(3+P_{d})
{P^{2}_{s}}}\\
\fl C^{12}_{xz}&=&
{\frac{{\boldsymbol\vert} \vec P(1)\times \vec P(2){\boldsymbol\vert}\ (P(2)^{2}-P(1)^{2})\cos\phi}{2(3+P_{d})
P^{2}_{s}}}\\
\fl C^{12}_{zz}&=&
{1\over 12}-{\frac{ P^{2}_{s}}{(3+P_{d})^{2}}}
+{\frac{P_{n}}{3
(3+P_{d})
P^{2}_{s}}}
\\
\fl C^{12}_{zy}&=&(P(1)^{2}-P(2)^{2}){{\boldsymbol\vert} \vec P(1)\times \vec P(2){\boldsymbol\vert}}\sin\phi\over{
2(3+P_{d})
P^{2}_{s}}\\
\fl C^{12}_{xy}&=&0\, , 
\end{eqnarray}  
where $P_{s}=|{\vec P}(1)+{\vec P}(2)|$ ,
$P_{n}= 4P(1)^{2}P(2)^{2}+2\vec P(1)\cdot \vec P(2)(P(1)^{2}+P(2)^{2})-\sin^{2}\theta$
and $P_{d}=\vec P(1)\cdot\vec P(2)$.
We have done a detailed graphical study of the correlations and squeezing for
various values of the independent parameters.  While the study reveals that
squeezing and correlations coexist and are equally more pronounced in certain
ranges, there are also narrow regions where one exists in the absence of the
other.  All these aspects are revealed in the figures 4-6.
\Figure
{Variation of spin-spin correlations 
$C_{xx}\ (+),C_{yy}\ (\bullet),C_{zz}\ (\times),
C_{xz}\ (\circ),C_{yz}\ (\star)$ and 
squeezing $Q\ (\Diamond)$ with respect to $\theta$, for
$P(1)=0.9,\ P(2)=0.85$ and $\phi=0^{\circ}$.}
\Figure{Variation of spin-spin correlations 
$C_{xx}\ (+),C_{yy}\ (\bullet),C_{zz}\ (\times),
C_{xz}\ (\circ),C_{yz}\ (\star)$ and 
squeezing $Q\ (\Diamond)$ with respect to $\theta$, for
$P(1)=0.95,\ P(2)=0.92$ and $\phi=5^{\circ}$.}
\Figure{Variation of spin-spin correlations 
$C_{xx}\ (+),C_{yy}\ (\bullet),C_{zz}\ (\times),
C_{xz}\ (\circ),C_{yz}\ (\star)$ and 
squeezing $Q\ (\Diamond)$ with respect to $\theta$, for
$P(1)=0.85,\ P(2)=0.95$ and $\phi=10^{\circ}$.}

It is therefore of interest to study more general 
cases of coupling of the sub systems in order to identify definite 
relationship between correlations and
squeezing.  In the context of quantum computation, the
nature of coupled states has been studied \cite{ch} under the following 
configurations:\\
(1) $\rho _{_{C}}$ is strongly separable ; i.e., $\rho _{_{C}}=\rho(1)\otimes\rho(2)$\\
(2) $\rho _{_{C}}$ is weakly separable ; i.e., $\rho _{_{C}}=\sum p_{i}\ \rho _{i}(1)
\otimes\rho _{i}(2)$, $\sum p_{i}=1$, $p_{i}\geq 0$\\
(3) $\rho _{_{C}}$ is non-separable ; i.e., $\rho _{_{C}}$ cannot be expressed
as in (1) and (2).

The third configuration is indeed recognized as possessing quantum entanglement.  
We have discussed the strongly separable mixed state case in this paper for the
particular case of $s_{1}={1\over 2}$ and $s_{2}={1\over 2}$. 
We wish to look at the squeezing and the correlation aspects for the cases (2)
and (3) in a sequel to this paper.    
               
\ack Two of us acknowledge with thanks the Council of Scientific and Industrial
Research (CSIR), India
for support through the award of Emeritus Scientistship to GR and
Senior Fellowship to PND.

\newpage
\begin{table}
\caption{Squeezed spin states specified by their non-zero spherical 
tensor parameters in LF and variances in $S_{x_{_{0}}}$
and $S_{y_{_{0}}}$.}
\label{tab}
\begin{indented}
\item[]\begin{tabular}{@{}lllllll}
\br
Spin value & $t^{2}_{0}$ & $t^{2}_{2}$ & $t^{1}_{0}$ & $\Delta S^{2}_{x_{_{0}}}$
& $\Delta S^{2}_{y_{_{0}}}$ & ${\textstyle{1\over 2}}\vert\langle S_{z_{_{0}}}\rangle\vert$\\
\mr
$3/2$ & $0.9$ & $0.3$ & $1.25$ & $1.17$ & $0.34$ & $0.7$\\   
$3/2$ & $0.7$ &$0.5$ & $1.06$ & $1.5$ & $0.28$ & $0.6$\\
$3/2$ & $0.61$ & $0.49$ & $0.99$ & $1.54$ & $0.34$ & $0.55$\\
$3/2$ & $0.41$ & $0.63$ & $0.81$ & $1.82$ & $0.27$ & $0.45$\\
$1$ & $0.7$ & $0.65$ & $0.8$ & $0.876$ & $0.12$ & $0.12$\\
$1$ & $0.5$ & $0.45$ & $0.9$ & $0.81$ & $0.28$ & $0.37$\\
$1$ & $0.4$ & $0.65$ & $0.5$ & $0.94$ & $0.197$ & $0.204$\\
$1$ & $0.3$ & $0.49$ & $0.7$ & $0.83$ & $0.27$ & $0.286$\\
\br
\end{tabular}
\end{indented}
\end{table}
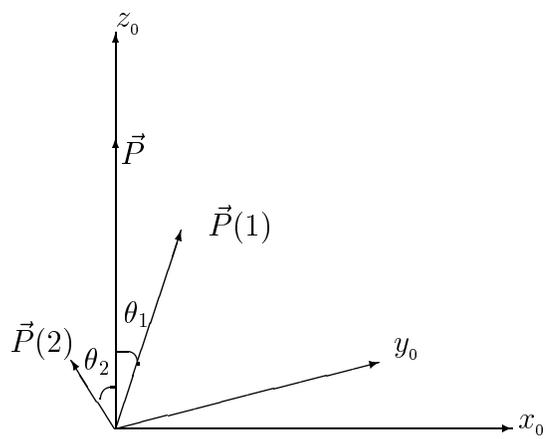
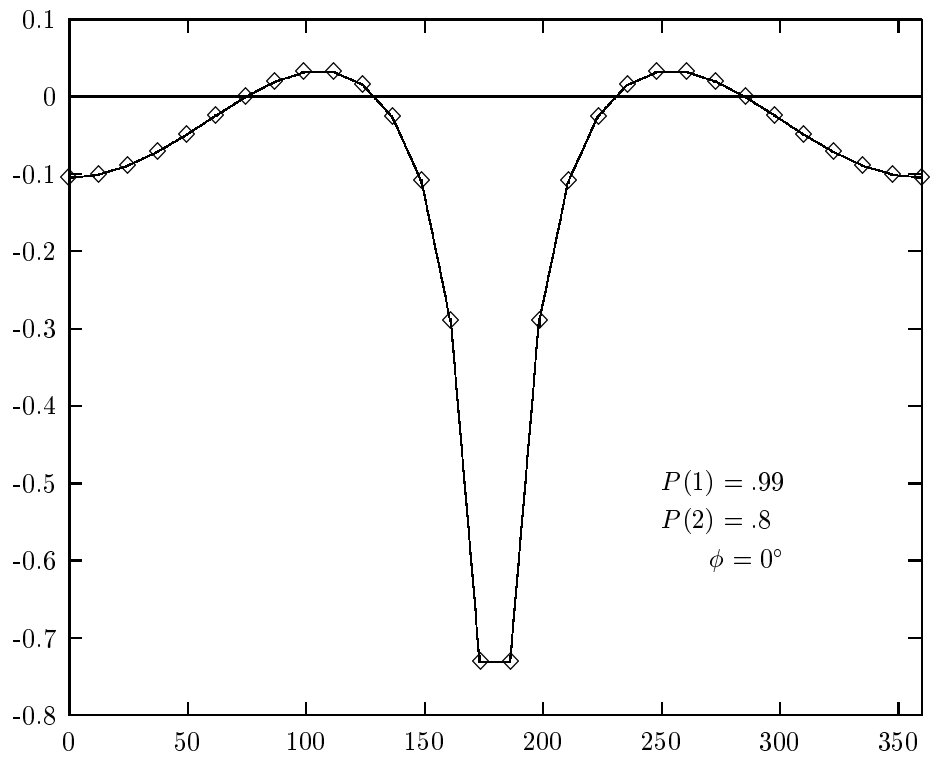
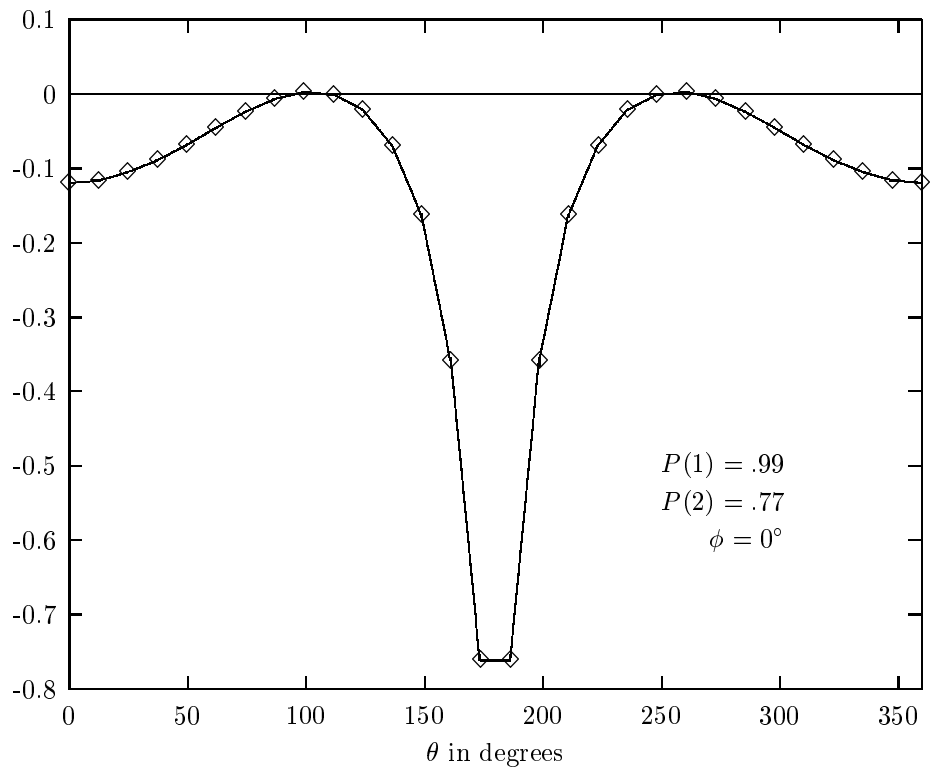
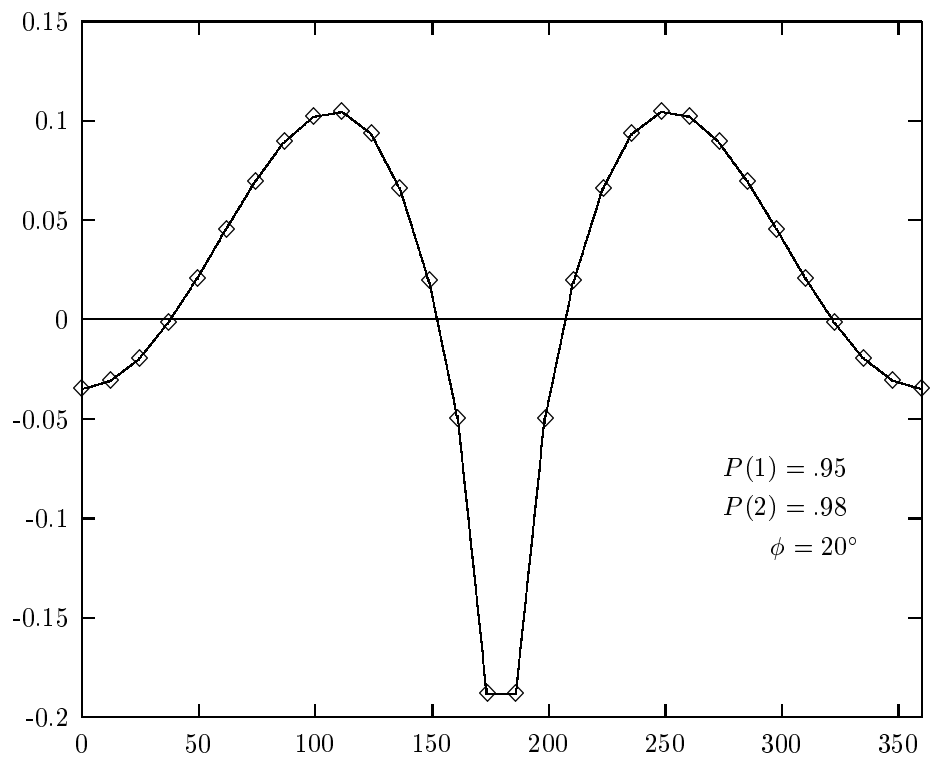
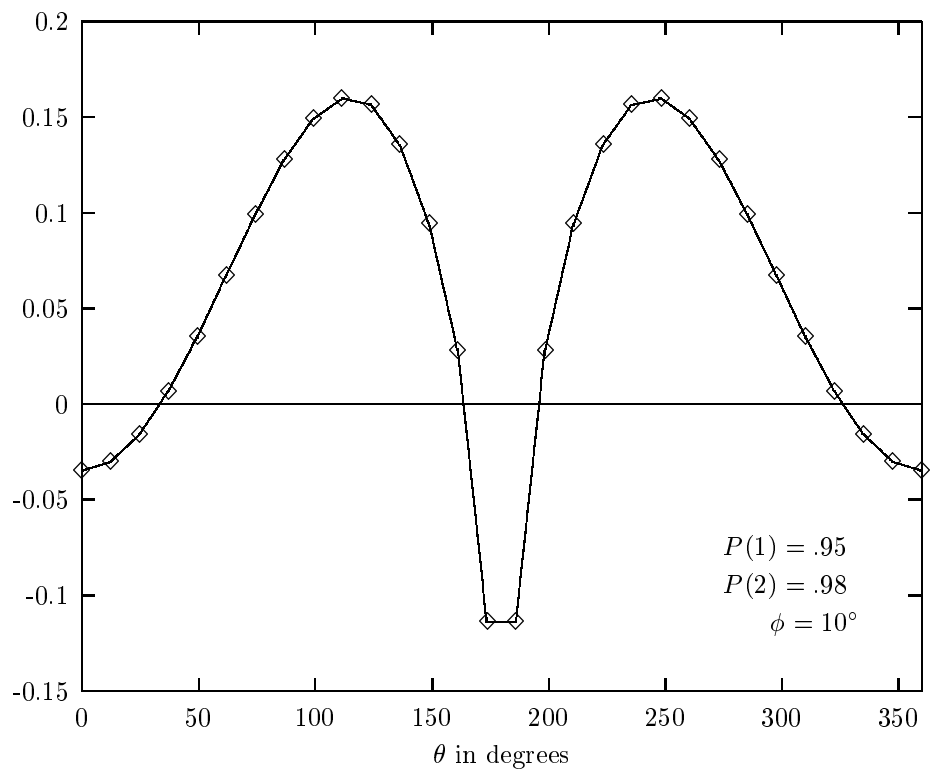
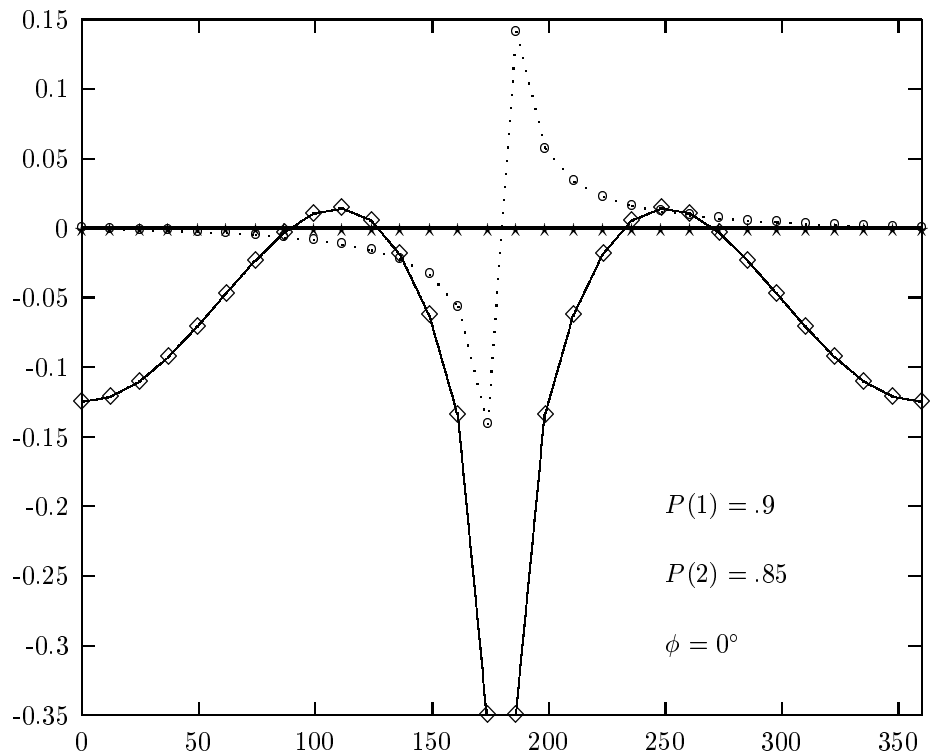
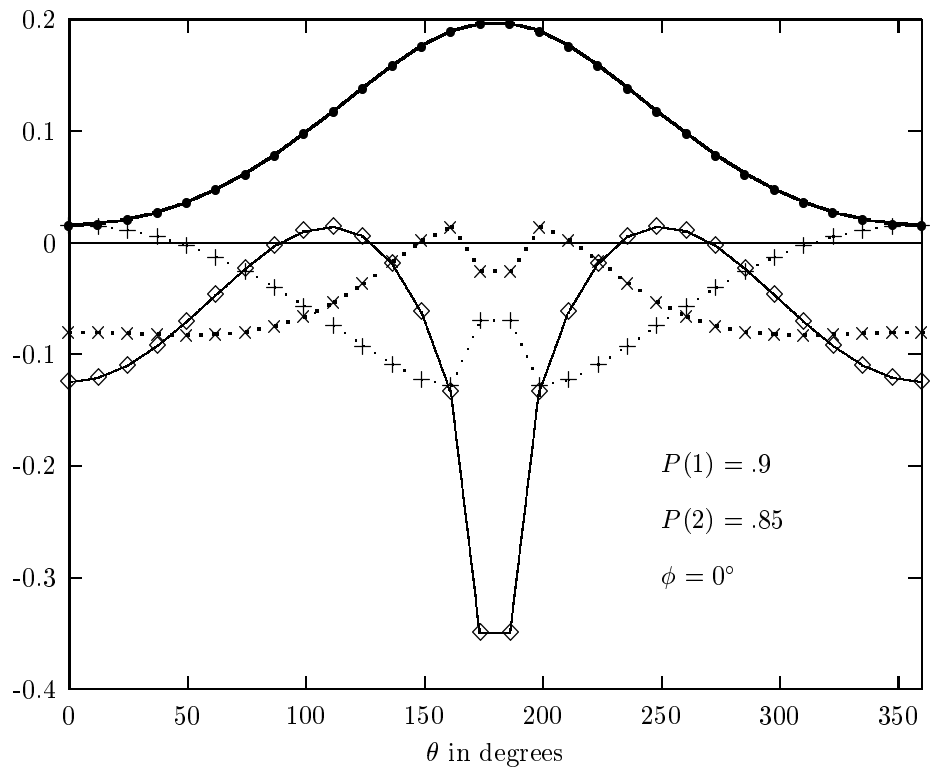
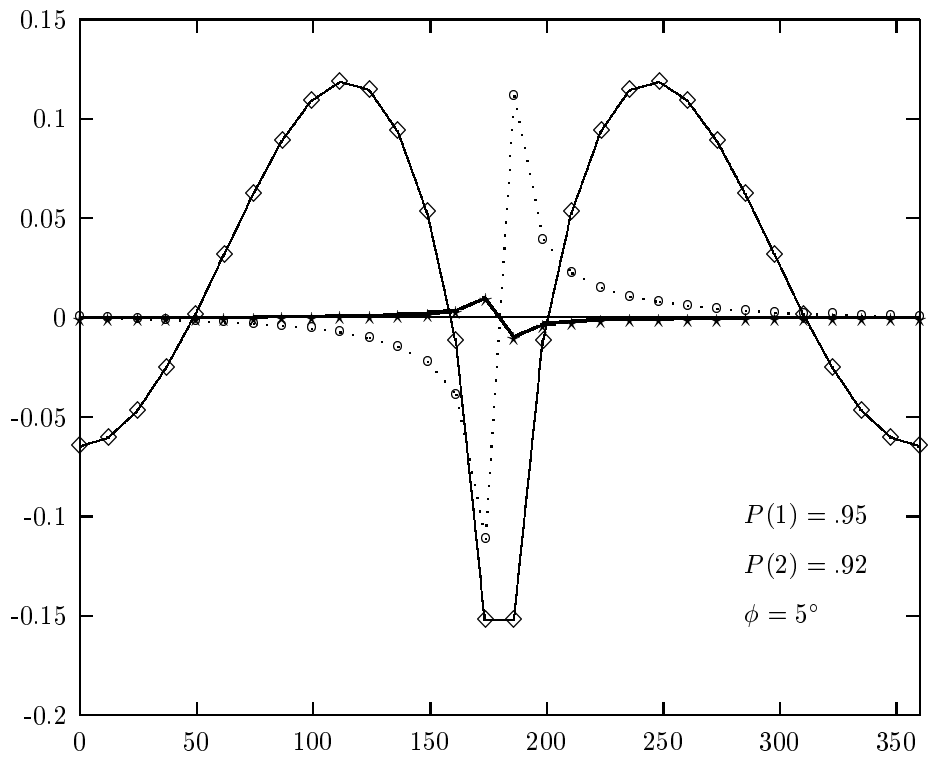
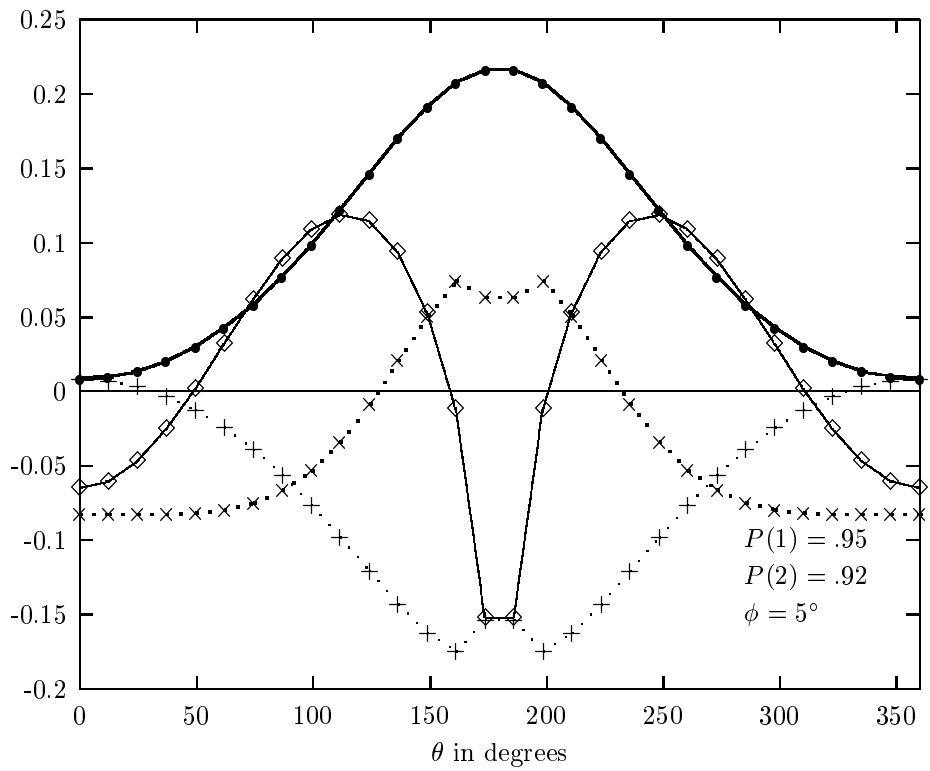
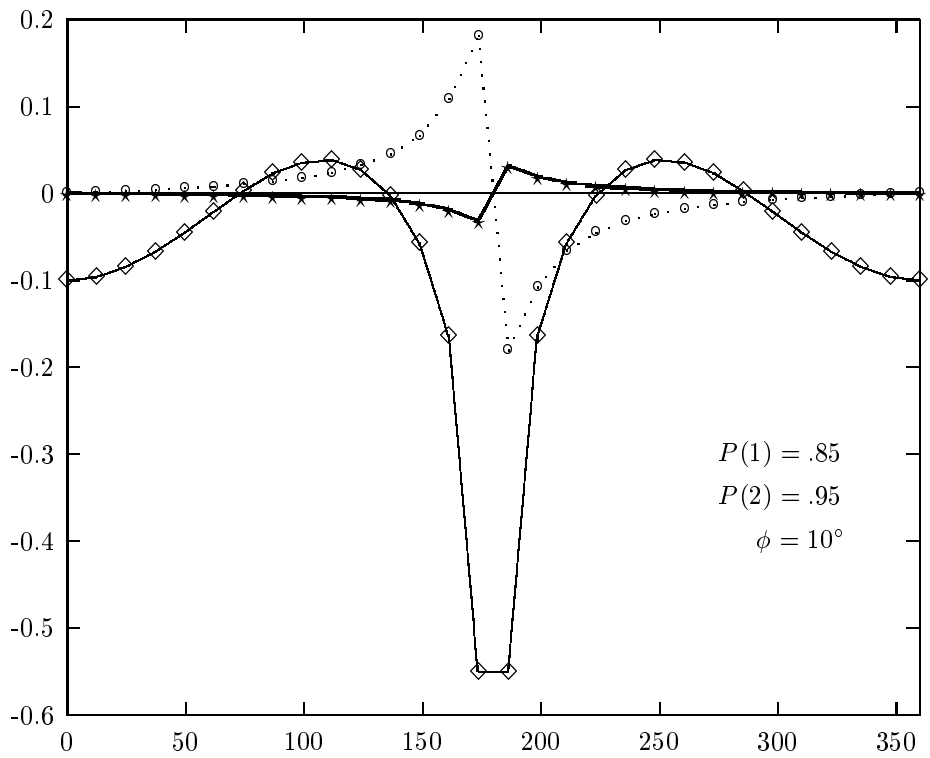
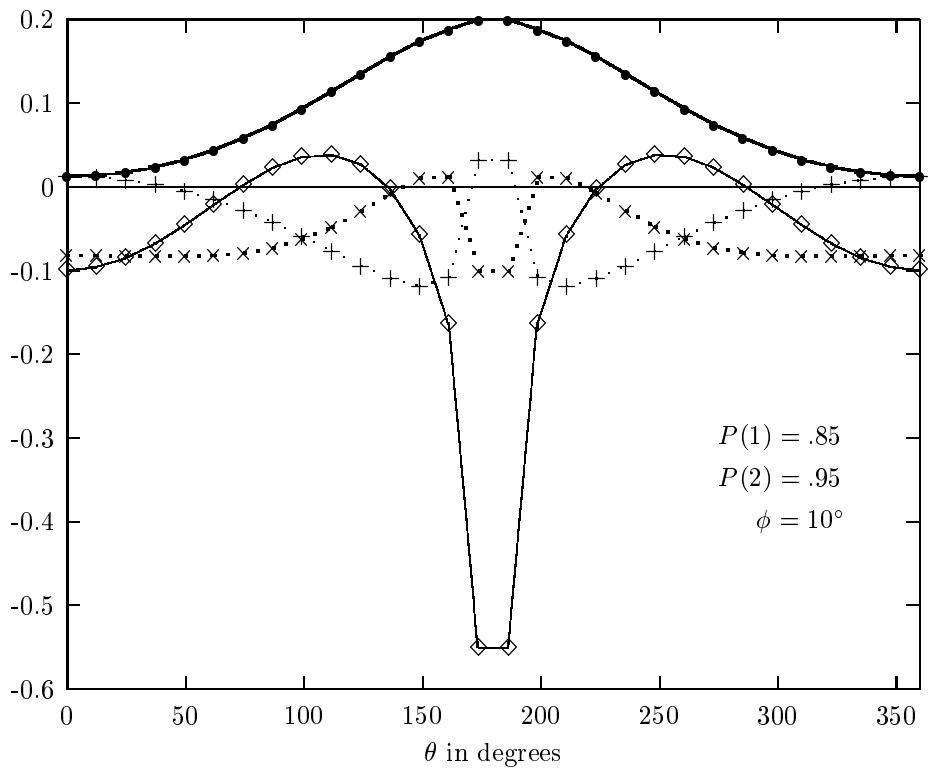
\end{document}